\documentclass[prd,preprint,tightenlines,floatfix,showpacs,preprintnumbers,nofootinbib,eqsecnum]{revtex4}

 \usepackage[dvips,final]{graphicx}
  \usepackage{amssymb}
   \usepackage{amsmath}
    \usepackage{amsfonts}
     \usepackage{epsfig}
      \usepackage{bm} % bold math

\begin{document}

\begin{flushright}
\vskip1cm
\end{flushright}

\title{Thermodynamics of accelerated fermion gas and instability at Unruh temperature}

\author{George Y. Prokhorov$^1$}
\email{prokhorov@theor.jinr.ru}

\author{Oleg V. Teryaev$^{1,2,3}$}
\email{teryaev@jinr.ru}

\author{Valentin I. Zakharov$^{2,4,5}$}
\email{vzakharov@itep.ru}
\affiliation{
{$^1$\sl
%Bogoliubov Laboratory of Theoretical Physics,
Joint Institute for Nuclear Research, 141980 Dubna,  Russia \\
$^2$ \sl
Institute of Theoretical and Experimental Physics, NRC Kurchatov Institute,
117218 Moscow, Russia
 \\
$^3$ \sl
M. V. Lomonosov Moscow State University, 117234 Moscow, Russia
 \\
$^4$ \sl
School of Biomedicine, Far Eastern Federal University, 690950 Vladivostok, Russia\\
$^5$ \sl
Moscow Institute of Physics and Technology, 141700 Dolgoprudny, Russia
}\vspace{1.5cm}
}

%%%%%%%%%%%%%%%%%%%%%%%%%%%%%%%%%%%%%%%%%%%%%%%%%%%%
\begin{abstract}
\vspace{0.5cm}
We demonstrate that the energy density of an accelerated fermion gas evaluated
within quantum statistical approach in Minkowski space is related to a quantum correction
to the vacuum expectation value of the energy-momentum tensor
in a space with non-trivial metric and conical singularity.
The key element of the derivation is the existence of a novel class of polynomial
Sommerfeld integrals.  The emerging duality of quantum statistical and geometrical approaches is explicitly checked  at temperatures $T$ above or
equal to the Unruh temperature $T_U$.
Treating the acceleration as an imaginary part of the chemical potential
allows for an analytical continuation to
 temperatures $T<T_U$. There is a discontinuity at $T=T_U$ manifested in
the second derivative of the energy density with respect to the temperature.
Moreover, energy density becomes negative
at $T<T_U$, apparently indicating some instability.
Obtained results might have phenomenological implications for the physics of heavy-ion collisions.
\end{abstract}
%%%%%%%%%%%%%%%%%%%%%%%%%%%%%%%%%%%%%%%%%%%%%%%%%%%%

\maketitle

%================
\section{Introduction}
\label{Sec:Intro}
%================

Study of collective quantum properties of relativistic matter is crucial
for the descriptions of media under extreme
conditions, in particular, of the quark-gluon plasma produced in heavy ion collisions.
It led to the discovery of new important chiral phenomena \cite{Kharzeev:2012ph}
such as chiral magnetic
effect, as well as
the influence of rotation and magnetic fields on polarization
\cite{Rogachevsky:2010ys, Becattini:2019ntv, Florkowski:2018fap, Baznat:2017jfj}
and the phase diagram   \cite{Jiang:2016wvv, Chernodub:2017ref, Wang:2018sur}.
A lot of efforts was made to improve our understanding
of the effects associated with rotation and magnetic fields, while the role of acceleration $a$ has been
much less discussed in this context.

%-----------------------------------------------------------------------------
Most recently, the situation has been changing due to development of a novel approach to
the quantum statistical physics based on the use of the Zubarev density operator. There exists now a systematic way to include the acceleration $a$ into the parameters characterizing equilibrium and evaluate perturbative expansion in the ratio $a/T$.
In particular, the energy densities of accelerated gas of massless particles with spins $0$ and $1/2$ were
evaluated explicitly \cite{Buzzegoli:2017cqy, Prokhorov:2019cik}.

%-----------------------------------------------------------------------------
As first noted in \cite{Becattini:2017ljh, Prokhorov:2019cik} the quantum statistical approach - rather unexpectedly - is sensitive to the Unruh temperature \cite{Unruh:1976db}
\begin{eqnarray}
T_U=\frac{a}{2\pi}\,.
\label{TU}
\end{eqnarray}
Let us remind that $T_U$ is the temperature of the radiation seen by an accelerated observer. Within the quantum statistical approach the energy density changes its sign at $ T=T_U $.

%-----------------------------------------------------------------------------
The Unruh effect is seen by an observer accelerated in Minkowski vacuum. In this case the relation (\ref{TU}) establishes a one-to-one correspondence between temperature and acceleration. The quantum statistical approach, on the other hand, treats $T$ and $a$ as independent parameters.
We borrow the interpretation of these states at $T > T_U$ from quantum field theory on the background of a space with horizon see e.g. \cite{Dowker:1994fi, Dowker:1987pk}. Namely, the Euclidean version of the Rindler space with a conical singularity provides an adequate image for a state with $T$ and $a$ being independent parameters.

%-----------------------------------------------------------------------------
As for the temperatures $T < T_U$, we argue that a proper analytical continuation of the energy density to this region can be worked out by using the Fermi distribution with the acceleration providing an imaginary part of the chemical potential \cite{Prokhorov:2019cik, Prokhorov:2018qhq}. To this end we need a non-perturbative representation for the energy density for the fermion gas at non-vanishing $(T,a)$. We did suggest such a representation to be valid in our previous paper \cite{Prokhorov:2019cik}. This integral representation was fitted to reproduce the first three terms of the perturbative expansion in $a/T$ and here we associate it with a novel type of {\it polynomial} Sommerfeld integrals which demonstrate, in the particular case of the energy density, the absence of perturbative terms beyond the first three terms known explicitly.

%-----------------------------------------------------------------------------
To summarize, at $T>T_U$ we have two {\it dual} representations for the energy density of the fermion gas. One is provided by the integral representation, see (\ref{main}), derived within the quantum statistical approach. The other one is given by the quantum correction to the vacuum expectation value of the $T_{00}$ component of the energy-momentum tensor in a space with horizon and conical singularity. The two representations, indeed, turn identical upon the proper identification of the corresponding parameters. With a stretch of imagination, we can say that starting with the thermodynamics of the accelerated gas we get the horizon emerging as a result of summation of the perturbative expansion in $a/T$. Also it is important to notice, that since the results, obtained in \cite{Dowker:1994fi, Dowker:1987pk}, are nonperturbative, and all the corrections above $ a^4 $ are equal to zero, we get one more evidence of polynomiality of energy density.

%-----------------------------------------------------------------------------
At temperatures $T<T_U$ the quantum statistical approach provides us with the means to evaluate the energy density of the fermion gas in the one-loop approximation. The explicit expression obtained in this way differs from the naive use of the (finite) perturbative series valid at $T>T_U$. There is a discontinuity at the point $T=T_U$ \cite{Prokhorov:2018qhq, Prokhorov:2019cik}. Analytic continuation allows to associate this instability at $ T=T_U $ with the crossing of the pole of the Fermi distribution in the complex
plane.
This pole is a nonperturbative manifestation
of the observation that acceleration appears
as an imaginary chemical potential \cite{Prokhorov:2018qhq}. However, the transition from $T>T_U$ to $T<T_U$ is rather smooth, so that only the second derivative from the energy density with respect to the temperature experiences a jump at $T=T_U$.

We interpret the behaviour of $\rho(T)$ around $T=T_U$ as indication of an instability. The negative sign of $\rho (T)$ at $T<T_U$ implies decay of the state into particles with positive energies (compensated by occupation of the corresponding levels with negative energy) and Minkowskian vacuum. An analogy to this process provided,  for example, by the superradiance from the ergosphere of a rotating black hole, where negative energy levels also exist \cite{Brito:2015oca}.

On the other hand, in the framework of the approach with space with a boundary, when $ T=T_U $ the conical singularity disappears and the cone turns into a plane.
So we see that this phenomenon is echoed by a quantum
instability which is arising at the same point.
It is amusing that a similar picture arises \cite{Pimentel:2018fuy, Gies:2015hia}
in the context of vacuum stability in external fields.

We also discuss the application of instability at Unruh temperature to the description of heavy-ion collisions. The pioneering attempts to relate the thermalization and the universality of the hadronization temperature  to the Unruh effect was made in Refs. \cite{Kharzeev:2006aj,Castorina:2007eb, Becattini:2008tx}.  The observation of the instability existence allows us to introduce the picture of hadronization which proceeds through the stage of formation of a state with high acceleration and temperature lower than $T_U$. The instability is then responsible for the decay of this state into final hadrons.

The paper has the following structure. Section \ref{Sec:energ den perturb}
discusses perturbative results for the energy density of
an accelerated massless fermion gas, obtained in the framework of the quantum-statistical approach. Section \ref{Sec:poly} demonstrates the possibility of representing a perturbative result in terms of Sommerfeld integrals of a new type and shows by integration in the complex plane that these integrals are polynomial.
Section \ref{Sec:dual} is devoted to quantum field theory with a conic singularity and shows that the results of this approach exactly coincide with the quantum-statistical approach. Section \ref{Sec:inst} discusses analytic continuation into region $ T<T_U $ and shows the existence of an instability. Section \ref{Sec:disc} considers various aspects of this instability and emphasizes a parallel with the decay of vacua in strong external fields. The physical interpretation of the instability and possible phenomenological applications are also discussed in this section.
The conclusion is given in the Section \ref{Sec:Concl}. Technical details
related to the calculation of the order of the derivative of the energy
density with instability, and instability at repeated crossing of the pole,
are included in the appendices \ref{Sec:order} and \ref{Sec:beyond}.

%=================================================
\section{Energy density of accelerated fermion gas}
\label{Sec:energ den perturb}
%=================================================

The properties of a medium in a state of global thermodynamic equilibrium can be described by the quantum-statistical Zubarev density operator of the form \cite{Zubarev, Weert, Buzzegoli:2017cqy, Becattini:2017ljh}
\begin{eqnarray}
\hat{\rho}=\frac{1}{Z}\exp\Big\{-\beta_{\mu}(x)\hat{P}^{\mu}
+\frac{1}{2}\varpi_{\mu\nu}\hat{J}^{\mu\nu}_x+\zeta \hat{Q}
\Big\} \,,
\label{rho global}
\end{eqnarray}
where $\hat{P}$ is the 4-momentum operator, $\hat{Q}$ is the charge operator, $\hat{J}_x$ are the generators of the Lorentz transformations displaced to the point $x$, and $ \varpi_{\mu\nu} $ is a tensor of thermal vorticity. Acceleration effects are contained in the term $ \varpi_{\mu\nu} \hat{J}^{\mu\nu}_x$, because
\begin{eqnarray}
\varpi_{\mu\nu} \hat{J}^{\mu\nu}_x=-2\alpha_{\mu}\hat{K}^{\mu}_x -2w_{\mu}\hat{J}^{\mu}_x \,,
\label{w}
\end{eqnarray}
where $ \alpha_{\mu}=a_{\mu}/T $ is the thermal acceleration, $ \hat{K}^{\mu}_x $ is the boost operator, $ w_{\mu}=\omega_{\mu}/T $ is the pseudovector of thermal vorticity, $ \hat{J}^{\mu}_x $ is the angular momentum operator. It is important to note that, as follows from (\ref{rho global}), from the point of view of quantum statistical mechanics, the effects of acceleration can be described in space with the usual Minkowski metric by adding a term with a boost to the density operator.

In \cite{Buzzegoli:2017cqy} a perturbation theory in $ \varpi $  was developed at a finite temperature. This perturbation theory was used in \cite{Prokhorov:2019cik} to calculate the mean value of the energy-momentum tensor of the accelerated fermion gas when $ w_{\mu}=0,\, \mu=0,\, m=0 $. The following expression was obtained for the energy density
\begin{eqnarray}
\rho = \frac{7 \pi ^2 T^4}{60}+\frac{T^2 |a|^2}{24} -\frac{17|a|^4}{960\pi^2}+\mathcal{O}(|a|^6)\,,
\label{perturb}
\end{eqnarray}
where $ |a|=\sqrt{-a^{\mu}a_{\mu}} $, in what follows we will denote $ |a|=a $.

It is easy to see that (\ref{perturb}) satisfies the condition
\begin{eqnarray}
\rho (T=\frac{a}{2\pi}) = 0\,,
\label{Unruh effect}
\end{eqnarray}
which is an indication of the Unruh effect from the point of view of the quantum-statistical approach with the Zubarev density operator \cite{Prokhorov:2019cik, Becattini:2017ljh, Florkowski:2018myy}.

%=================================================
\section{Novel class of polynomial Sommerfeld integrals}
\label{Sec:poly}
%=================================================

In this section, we discuss an interesting property related to the solution (\ref{perturb}) and the possibility of representing it in the form of a new type of Sommerfeld integrals and show that these integrals are polynomial.

In \cite{Prokhorov:2019cik} an integral representation was proposed for (\ref{perturb})
\begin{eqnarray}
\rho =  2 \int \frac{d^3 p}{(2\pi)^3}\Big(\frac{|\bold{p}|+ia}{1+e^{\frac{|\bold{p}|}{T}+\frac{ia}{2T}}}+\frac{|\bold{p}|-ia}{1+e^{\frac{|\bold{p}|}{T}-\frac{ia}{2T}}}\Big)+4 \int \frac{d^3 p}{(2\pi)^3}\frac{|\bold{p}|}{e^{\frac{2\pi |\bold{p}|}{a}}-1}\,.
\label{main}
\end{eqnarray}
It was shown in \cite{Prokhorov:2019cik}, that for $T>T_U$ the Eq.~(\ref{main}) exactly coincides with the perturbative result (\ref{perturb}) (with $ \mathcal{O}(|a|^6)=0 $).

Eq.~(\ref{main}) also receives further support from the consideration of the Wigner function \cite{Becattini:2013fla} on the basis of which, in particular, it is also possible to show the addition of an imaginary term with acceleration $\pm \frac{i a}{2}$  to the chemical potential \cite{Prokhorov:2019cik}. Eq.~(\ref{main}) is remarkable in that it automatically leads to the condition (\ref{Unruh effect}) since in this case the bosonic "counter-term" in (\ref{main}) turns out to be exactly equal to the first integral with the opposite sign. As discussed in   \cite{Prokhorov:2019cik}, this fact is a manifestation of the Unruh effect within quantum statistical mechanics   \cite{Becattini:2017ljh}. Indeed, the energy density of the Minkowskian vacuum is normalised to zero, and (\ref{Unruh effect}) demonstrates that the energy density vanishes at the Unruh temperature.

The most unusual property of the Eq.~(\ref{main}) is the appearance of a bosonic type contribution in it. This contribution corresponds formally to a gas of massless bosons with 4 degrees of freedom and in limit $T\to 0$ it is the only non-vanishing contribution
\begin{eqnarray}
\rho (T\to 0)= 4 \int \frac{d^3 p}{(2\pi)^3}\frac{|\bold{p}|}{e^{\frac{2\pi |\bold{p}|}{a}}-1}\,.
\label{Rindler}
\end{eqnarray}
The appearance of such a term can be qualitatively related to the equivalence principle \cite{Fulling:2018lez}. Somewhat similar counter-term was also introduced in   \cite{Stone:2018zel, Becattini:2017ljh}, while its bosonic nature can be attributed to imaginary chemical potential, connected with acceleration. Note also, that our counter-term is positive, while in \cite{Becattini:2017ljh} a similar counter-term is negative.

From a mathematical point of view, integrals of the form (\ref{main}) are a new type of Sommerfeld integrals (look, e.g. \cite{Stone:2018zel}). Similar Sommerfeld integrals have already been discussed in the literature in various contexts, however (\ref{main}) differs by the presence of an imaginary term in the exponent.

A remarkable property of the integrals (\ref{main}) is their polynomiality.
Here we present a simple method that allows us to show this polynomiality and better understand its source. We will use the Blankenbecler's method   \cite{Blankenbecler, Sprung, Burov}, originally used in nuclear physics. We generalize this method to the case of antisymmetric weight function and imaginary chemical potential. Eq.~(\ref{main}) can be converted to
\begin{eqnarray}
\rho &=&  \frac{a^4}{120\pi^2}+\frac{T^4}{\pi^2}\Big(\Big[\int_0^{\infty}\frac{x^3 dx}{e^{x+iy}+1}+\int_0^{\infty}\frac{x^3 dx}{e^{x-iy}+1}\Big] \nonumber \\
&&+2iy \Big[\int_0^{\infty}\frac{x^2 dx}{e^{x+iy}+1}-\int_0^{\infty}\frac{x^2 dx}{e^{x-iy}+1}\Big]\Big)
\,,
\label{rho with Is1 Is2}
\end{eqnarray}
where $y=\frac{a}{2T}$ and we substituted the value of a Bose integral right away, as its polynomiality is obvious in advance. Let us consider a more general case of integrals of the type  (\ref{rho with Is1 Is2}), for almost arbitrary weight function. Two types of integrals are possible:
\begin{eqnarray}
I_{s1} &=& \int_0^{\infty}\frac{f(x) dx}{e^{x+iy}+1}+\int_0^{\infty}\frac{f(x) dx}{e^{x-iy}+1}\,, \nonumber \\
I_{s2} &=& \int_0^{\infty}\frac{g(x) dx}{e^{x+iy}+1}-\int_0^{\infty}\frac{g(x) dx}{e^{x-iy}+1}
\,,
\label{Is1Is2}
\end{eqnarray}
where $f(x)=-f(-x)$ is an odd function, and $g(x)=g(-x)$  is an even function. In the case of  (\ref{rho with Is1 Is2}) $f(x)=x^3$ and $g(x)=x^2$. First we calculate the integral $I_{s1}$ and assume that $0<y<\pi$. To do this, we integrate both terms by parts in  (\ref{Is1Is2}) and make the change of variables
\begin{eqnarray}
I_{s1} &=& -\int_0^{\infty}F(x) S(x+iy) dx-\int_0^{\infty}F(x) S(x-iy) dx\,,\nonumber \\
S(x+iy) &=& \frac{\partial}{\partial x}\frac{1}{e^{x+iy}+1}=-\frac{e^{x+iy}}{(e^{x+iy}+1)^2}
\,,\nonumber \\
F(x) &=& \int_0^x f(x) dx\,.
\label{Is1 calc1}
\end{eqnarray}
After the change of variables and using the parity $S(x)=S(-x)$ and $F(x)=F(-x)$, we get
\begin{eqnarray}
I_{s1} &=& - \int_{iy-\infty}^{iy+\infty}F(x-iy) S(x) dx\,.
\label{Is1 calc2}
\end{eqnarray}
Note that for obtaining  (\ref{Is1 calc2}), the oddness of function $f(x)$, is crucial, leading to an even function $F(x)$. Also the presence of two integrals with $+iy$ and $-iy$  (which correspond to the appearance of contributions with $+ia$ and $-ia$ in  (\ref{main})) is significant.

\begin{figure*}[!h]
\begin{minipage}{0.45\textwidth}
 \centerline{\includegraphics[width=1\textwidth]{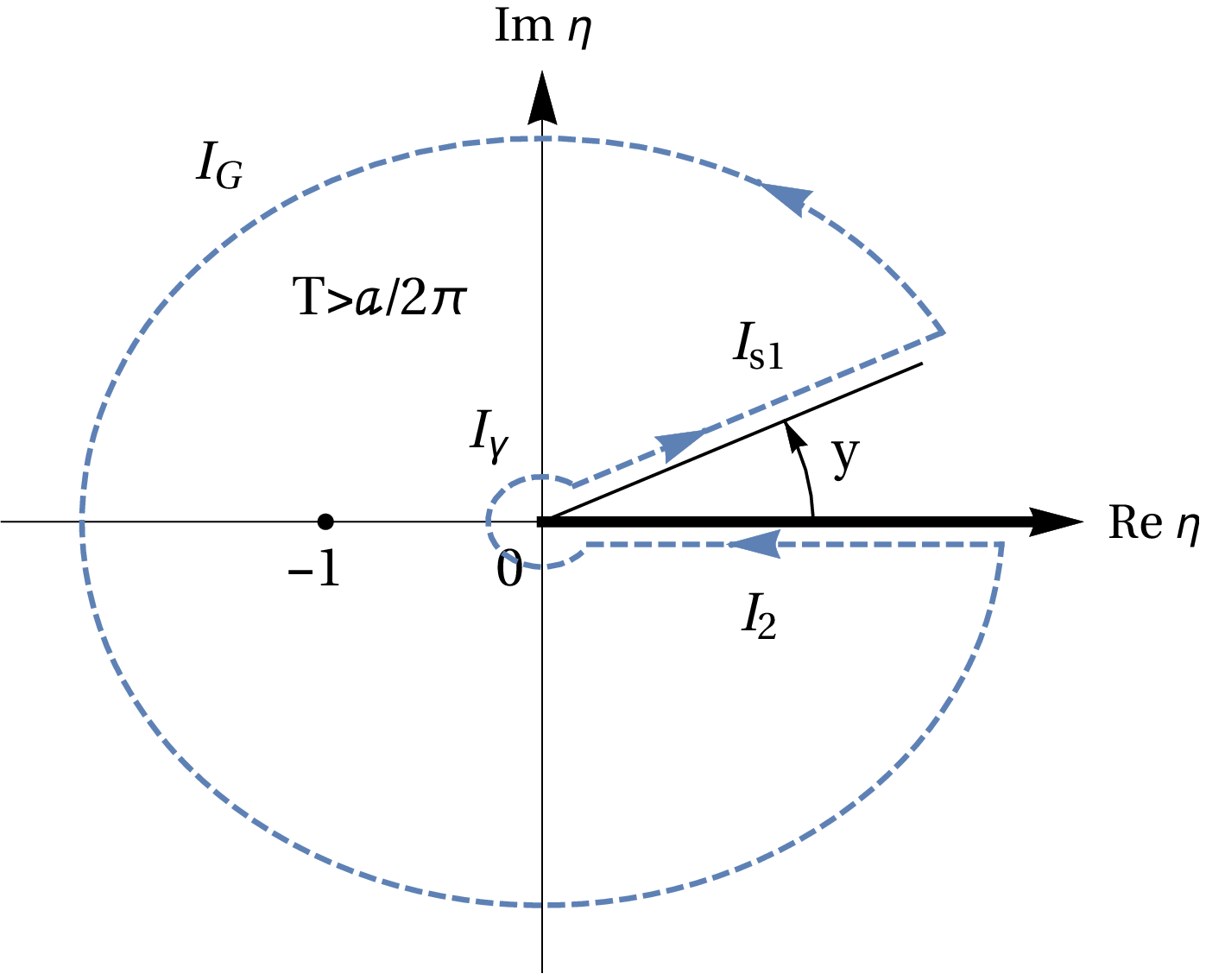}}
\end{minipage}\hspace{0.7 cm}
\begin{minipage}{0.45\textwidth}
 \centerline{\includegraphics[width=1\textwidth]{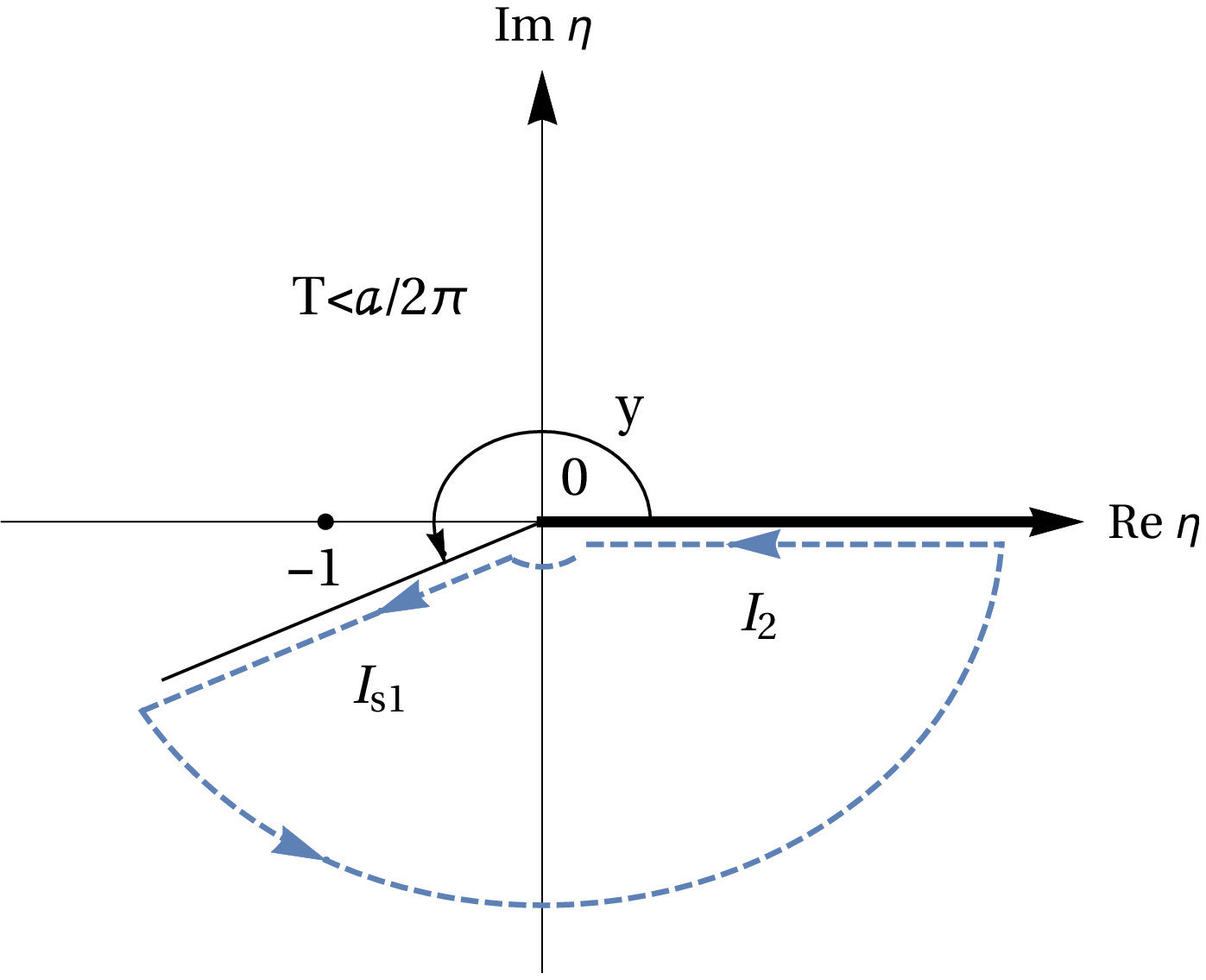}}
\end{minipage}\vspace{1.5 cm}
\caption{The slope of the integration contour $y=\frac{a}{2T}$ is determined by the acceleration. \textit{Left}: integration contour in  (\ref{Is1 int eta}) when $0<y<\pi$. \textit{Right}: integration contour in  (\ref{Is1 int eta}) when $y>\pi$.}
\label{fig:contour}
\end{figure*}

Now let us present the Taylor expansion of the function $F(x-iy)$, in the form of an exponent with the derivative (in other words, we use the translation operator)
\begin{eqnarray}
F(x-iy)=e^{x D}F(-iy)\,,\qquad D=\frac{\partial}{\partial(-iy)}\,.
\label{transl}
\end{eqnarray}
Making change of variables, $\eta=e^{x}$, we get
\begin{eqnarray}
I_{s1} &=& \int_{I_{s1}}\frac{\eta^D d\eta}{(\eta+1)^2} F(-iy)\,,
\label{Is1 int eta}
\end{eqnarray}
where integration is along a straight line in the complex plane at an angle $y=\frac{a}{2T}$ to the real positive semi-axis (the left plot in Fig.~\ref{fig:contour}). Non-zero slope of the integration contour with respect to the positive semi-axis is a direct consequence of existence of the imaginary chemical potential, and this distinguishes our calculation from a similar one in \cite{Blankenbecler, Sprung, Burov}.
The integrand in  (\ref{Is1 int eta}) has a second-order "Regge-like" pole \cite{Gribov:2009zz} in $\eta=-1$, stemming from the Fermi distribution, and a cut along the positive real semi-axis. To calculate the integral $I_{s1}$, one has to close the contour in the complex plane at infinity as shown in the left plot in Fig.~\ref{fig:contour}.

The integral $I_G=I_{\gamma}=0$, since following   \cite{Blankenbecler, Sprung} we assume $|D|<1$. The integral along the whole contour $I_C$ is then equal to
\begin{eqnarray}
I_C=I_{s1} + I_2\,.
\label{Im plane sum}
\end{eqnarray}
At the same time, due to the rotation angle equal to $2 \pi$ in the case of $I_2$
\begin{eqnarray}
 I_2=-e^{2\pi i D}I_{s1}\,.
\label{I2 Is1}
\end{eqnarray}
Using the residue theorem, we get
\begin{eqnarray}
I_C &=& 2\pi i\, \mathrm{Res}_{\eta=-1}\frac{\eta^D}{(\eta+1)^2} F(-iy)\,,
\label{residue1}
\end{eqnarray}
and after calculating the residue at the second-order pole, we obtain
\begin{eqnarray}
I_{s1} &=& \frac{\pi D}{\sin(\pi D)} F(-iy)\qquad (0<y<\pi)\,.
\label{Is1 final}
\end{eqnarray}
Note that though the case considered is somewhat different from   \cite{Blankenbecler, Sprung} , where the keyhole contour was considered, the final formula is the same (differences will appear after crossing the pole, as we discuss below).

In the same way, it is possible to obtain similar expressions for $I_{s2}$
\begin{eqnarray}
I_{s2} &=& \frac{\pi D}{\sin(\pi D)} G(-iy)\qquad (0<y<\pi)\,, \nonumber \\
G(x) &=& \int_0^x g(x) dx\,.
\label{Is2 final}
\end{eqnarray}
To get a final answer, one needs to expand the function $\frac{\pi D}{\sin(\pi D)}$ into a Taylor series. We give the first four terms of the series
\begin{eqnarray}
\frac{\pi D}{\sin(\pi D)}=1+\frac{(\pi D)^2}{6}+\frac{7(\pi D)^4}{960}+\mathcal{O}\big((\pi D)^6\big)\,.
\label{pdsin Taylor}
\end{eqnarray}
Thus, the polynomiality of the energy is guaranteed by the polynomiality of the functions $F$ and $G$, or $f$ and $g$. Then the energy density in  (\ref{rho with Is1 Is2}) becomes
\begin{eqnarray}
\rho &=&  \frac{a^4}{120\pi^2}+\frac{T^4}{\pi^2}\Big(
\frac{\pi D}{\sin(\pi D)}\frac{(-iy)^4}{4}+2iy\frac{\pi D}{\sin(\pi D)}\frac{(-iy)^3}{3}
\Big)\Big|_{y=\frac{a}{2T}} \nonumber \\
&=&  \frac{7 \pi ^2 T^4}{60}+\frac{T^2 a^2}{24} -\frac{17a^4}{960\pi^2}
\qquad T>\frac{a}{2\pi}\,,
\label{pdsin poly}
\end{eqnarray}
where the condition $0<y<\pi$, necessary for the contour not to cross the pole,
 leads to the condition that the
temperature be higher than the Unruh temperature $T>\frac{a}{2\pi}$. To summarize, it is essential for polynomiality  that Fermi distributions in
(\ref{main}) are taken with polynomial weights, and also that symmetric
combinations of integrals with $\pm i a$ appear (the odd weight function
must correspond to the sum of the integrals with $\pm i a$, and even - to their difference).
If we use the physical interpretation of \cite{Prokhorov:2019cik},
then only the total contribution of the modes with imaginary chemical potential
$+ i a$ and $- i a$ is polynomial.

%=================================================
\section{Duality of quantum statistical mechanics and quantum field theory in a space with boundary}
\label{Sec:dual}
%=================================================

In this subsection, we show that the energy density of an accelerated gas can also be calculated
in another way in the framework of field theory in a space with a conical singularity
\cite{Dowker:1994fi, Dowker:1987pk}.
Thus, we demonstrate duality between quantum statistical calculations and quantum field theory in a space with a boundary.

Consider the Rindler metric in the form
\begin{eqnarray}
ds^2=-r^2 d \eta^2 +d r^2 +d\bold{x}^2_{\perp}\,,
\label{rindler}
\end{eqnarray}
where $ \eta=\gamma \lambda,\, x=r\, \mathrm{cosh}\, \eta ,\,  t=r\, \mathrm{sinh}\, \eta$, and $ \gamma $ is a positive constant (in (\ref{rindler}) for convenience, unlike the rest of the text, we consider
definition of the metric such that $ g_{00}<0 $). The world lines with $ r=\mathrm{const},\, \bold{x}_{\perp}=\mathrm{const} $ correspond to uniformly accelerated motion. The relations between the proper acceleration $ a $ and the proper time $ \tau $ along these world lines, with the variables $ \lambda $ and $ r $ is determined by the formulas
\begin{eqnarray}
a=r^{-1},\quad \tau= \gamma r \lambda\,.
\label{rindler coord}
\end{eqnarray}
In particular, for a world line with $ r=\gamma^{-1} $ the proper acceleration is $ a=\gamma $, and the proper time is $ \tau=\lambda $. We emphasize, however, that one should not confuse constant $ \gamma $ with the proper acceleration $ a $, and the variable $ \lambda $ with the proper time $ \tau $.

When constructing the field theory at finite temperatures, it is necessary to consider the proper time $ \tau $ as a coordinate periodic in the inverse proper temperature $ T^{-1} $ and, therefore,  to identify the points $ \tau=\gamma r \lambda =0 $ and $ \tau=\gamma r \lambda =T^{-1} $. Accordingly, we need to identify $ \eta =0 $ and $ \eta = 1/(rT) $. According to  (\ref{rindler coord}), $ 1/(rT)=a/T$, and this ratio is a spatio-temporal constant \cite{Becattini:2017ljh}.

\begin{figure*}[!h]
\begin{minipage}{0.25\textwidth}
 \centerline{\includegraphics[width=1\textwidth]{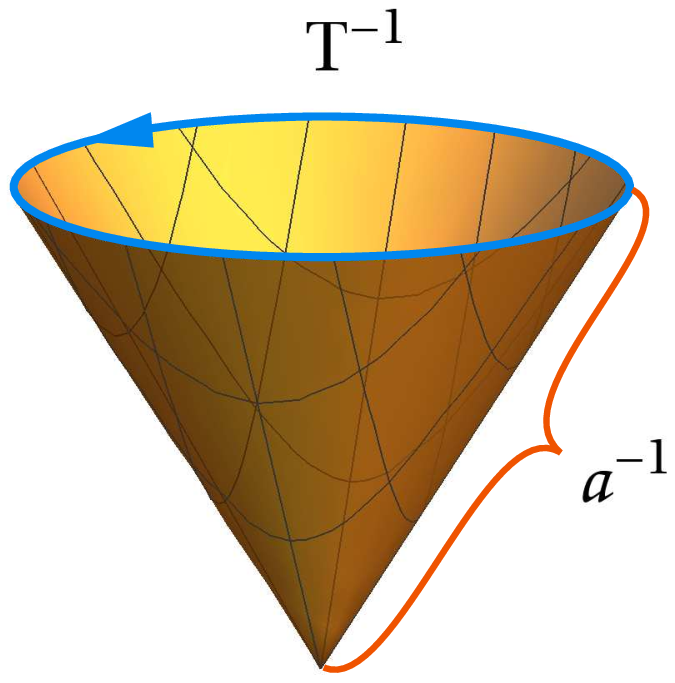}}
\end{minipage}\hspace{0.7 cm}
\begin{minipage}{0.45\textwidth}
 \centerline{\includegraphics[width=1\textwidth]{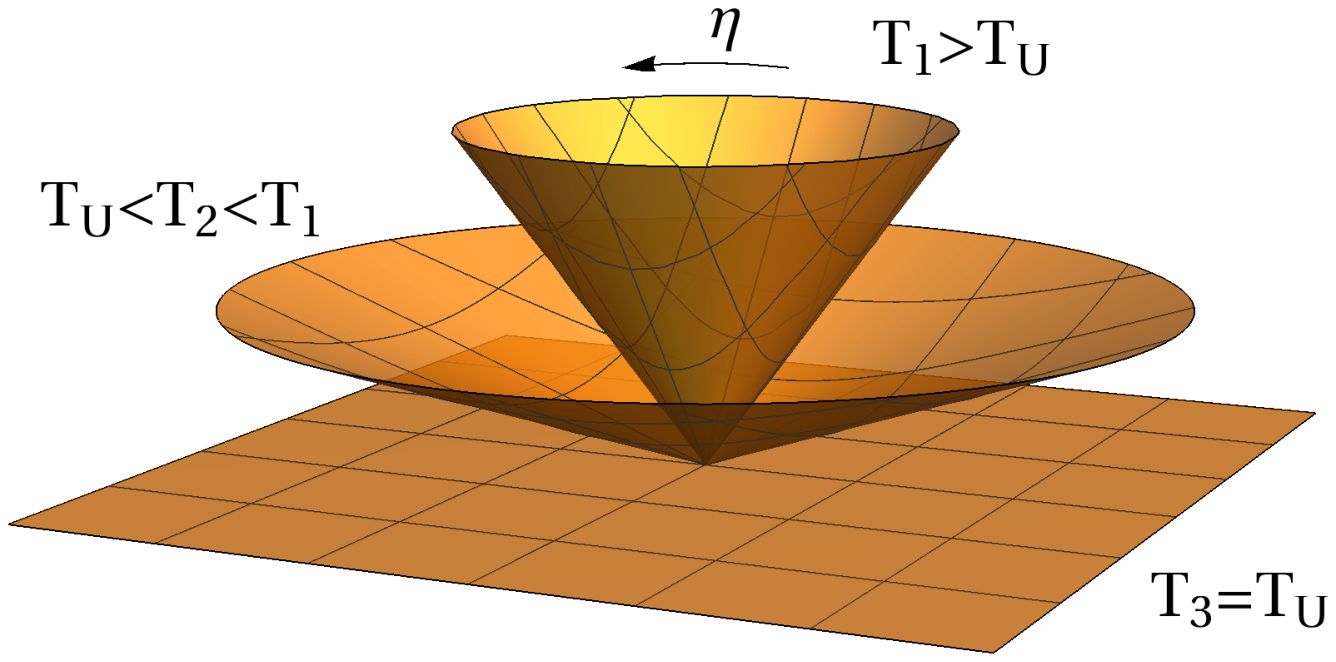}}
\end{minipage}\vspace{1.5 cm}
\caption{\textit{Left}: space with a conical singularity and the relationship of geometric characteristics, circumference and distance to the cone point, with statistical parameters, acceleration and temperature. \textit{Right}: transformation of a cone into a plane with the temperature decreasing to $ T= T_U $.}
\label{fig:cone}
\end{figure*}

Thus, the metric (\ref{rindler}), when considering field theory at finite temperatures, takes the form
\begin{eqnarray}
ds^2=r^2 d \eta^2 +d r^2 +d\bold{x}^2_{\perp}\,.
\label{rindler1}
\end{eqnarray}
Eq.~(\ref{rindler1}) describes the space, containing a flat two-dimensional
cone with an angular deficit $2\pi- a/T $. The world line of a uniformly accelerated object corresponds to a circle on the cone. Moreover, according to (\ref{rindler coord}), the distance from the top of the cone to the circle determines the inverse acceleration for this world line, and the length of the given circle determines the inverse proper temperature, as it is shown on the left panel of Fig.~\ref{fig:cone}. An essential property of the
metric (\ref{rindler1}) is the presence of a conical singularity at $ r=0 $.
Note that, as in the Fig. \ref{fig:contour}, the angular deficit of the cone is determined by the ratio of acceleration to temperature, $ a/T $.

The quantum field theory in the space (\ref{rindler1}) can be constructed using
the heat kernel method \cite{Dowker:1994fi, Dowker:1987pk, Fursaev:1993qk}.
As a result, nonperturbative mean values for different operators can be obtained.
In particular, in \cite{Dowker:1994fi, Dowker:1987pk} the mean value for
the energy density of Weyl spinor field was calculated
\begin{eqnarray}
\rho=\frac{1}{2} \Big(\frac{7\pi^2 T^4}{60}+\frac{T^2}{24 r^2}-\frac{17}{960 \pi^2 r^4}\Big)\,.
\label{energ Dow}
\end{eqnarray}
In this case, the last term, which is independent of temperature, is associated with the vacuum energy arising due to the Casimir effect in the space with a horizon \cite{Mertens:2015ola}.

An amazing observation is that taking into account (\ref{rindler coord}),  we get from (\ref{energ Dow})
\begin{eqnarray}
\rho=\frac{1}{2} \Big(\frac{7\pi^2 T^4}{60}+\frac{T^2 a^2}{24}-\frac{17 a^4}{960 \pi^2}\Big)\,,
\label{energ Dow1}
\end{eqnarray}
and thus, there is a complete agreement with the result (\ref{perturb}) (the difference in the coefficient $ 1/2 $ is associated with half the number of degrees of freedom of the Weyl spinors in comparison with the Dirac spinors). This means that the energy density of accelerated matter can be calculated in two completely different ways: either by means of the statistical Zubarev density operator (\ref{rho global}) and calculation of corrections in flat space, or by considering space with boundary, which transforms to the space with the conical singularity (\ref{rindler1}) in the framework of the heat kernel approach.

The results obtained in the framework of this approach are nonperturbative, which corresponds to the nonperturbative nature of the heat kernel, which takes into account all the orders in $ a $. In particular, the expression (\ref{energ Dow1}) is an exact nonperturbative formula and higher-order corrections are absent at least when $ T>T_U $.

Thus, the polynomiality of (\ref{perturb}), which, as it was shown in the previous section, is connected with the polynomiality of Sommerfeld integrals,  is justified in the framework of the approach with conical singularity on the field theoretical side.

One could expect that the polynomiality of energy density and of other observables is related to the quantum anomalies. It is well known that quantum anomalies can lead to the suppression of higher order quantum corrections, and as the result, the exact expression of a physical quantity is described by the first terms of the quantum-field perturbative series. This statement is known as the Adler-Bardeen
theorem \cite{Adler:1969er} and is well known in quantum field theory.

Recently, it has been shown that anomalies play crucial role in hydrodynamics. In various contexts, the relationship of quantum anomalies with the chiral vortical, magnetic, and other chiral effects, as well as with the Hawking effect, was shown  \cite{Son:2009tf, Sadofyev:2010is, Stone:2018zel}. We would expect that quantum anomalies in hydrodynamics also should guarantee through a kind of the Adler-Bardeen
theorem, the polynomiality in acceleration. However, finding the proof of this statement remains an interesting unsolved problem.

%=================================================
\section{Instability at Unruh temperature}
\label{Sec:inst}
%=================================================

\subsection{Analytical continuation to the region $ T<T_U $}

From the point of view of the quantum-field approach described in the previous section, the angular deficit $ 2\pi-a/T $ cannot be less than $ 0 $ and therefore, the temperature satisfies the condition
\begin{eqnarray}
T\geq T_U\,.
\label{TgTU}
\end{eqnarray}
At $ T=T_U $, the cone turns into a plane as it is shown on the right panel of the Fig.~\ref{fig:cone} and the quantum-field approach with a conical singularity in its standard form does not allow us to study the region $ T<T_U $.

However, from the point of view of the quantum-statistical approach, we can consider this region. In particular, Eq.~(\ref{main}) describes the analytic continuation into the region $ T<T_U $, and the point $ T=T_U $ itself corresponds to the crossing of the pole $ \eta=-1 $, as it is shown on the Fig.~\ref{fig:contour}. As will now be shown, in the region $ T<T_U $ the perturbative results (\ref{perturb}) are inapplicable, and nonperturbative effects appear.

Consider (\ref{main}) in the domain $ T<a/2\pi $. At $0<y<\pi$ we have $I_C=0$ from the residue theorem and from  (\ref{Im plane sum}) we get
\begin{eqnarray}
0=I_{s1} + I_2\,.
\label{contour new}
\end{eqnarray}
But $I_2$ remains the same as in in the region $ T>a/2\pi $. It is therefore easy to get
\begin{eqnarray}
I_{s1} &=& e^{2\pi i D}\frac{\pi D}{\sin(\pi D)} F(-iy)=\Big(1+2 i \pi D-\frac{11(\pi D)^2}{6}-i (\pi D)^3 \nonumber \\
&&+\frac{127(\pi D)^4}{360}+\mathcal{O}\big((\pi D)^5\big)\Big)F(-iy)\,,
\label{rot Is1 final}
\end{eqnarray}
and the same formula for $I_{s2}$ with $G(-iy)$. Accordingly, for the energy density we get
\begin{eqnarray}
\rho &=&\frac{127 \pi ^2 T^4}{60}-\frac{11 T^2 a^2}{24} -\frac{17a^4}{960\pi^2}-\pi T^3 a+ \frac{T a^3}{4\pi}\,.
\label{en 2 domain}
\end{eqnarray}
We note an interesting fact - according to  (\ref{en 2 domain}) at $T<T_U$, the terms of odd powers in $a$ appear (there are no contradictions from the point of view of parity and Lorentz invariance, since in  (\ref{main}) acceleration appears as $a=\sqrt{-a_{\mu}a^{\mu}}$).

One can see that nonperturbative result (\ref{en 2 domain}) is different from the perturbative one (\ref{perturb}).

\subsection{Instability}

The analytic continuation to the region $ T<T_U $, considered in the previous subsection, allows us to show the existence of instability at the Unruh temperature.

Consider the energy density in two regions $T>T_U$ and $T<T_U$. $\rho_{T>T_U}$  is given by  (\ref{pdsin poly}), and $\rho_{T<T_U}$ is given by  (\ref{en 2 domain}). It is easy to show that
\begin{eqnarray}
\rho_{T>T_U}(T\to T_U)&=&\rho_{T<T_U}(T\to T_U)=0\,, \nonumber \\
\frac{\partial}{\partial T}\rho_{T>T_U}(T\to T_U)&=&\frac{\partial}{\partial T}\rho_{T<T_U}(T\to T_U)=\frac{a^3}{10\pi}\,, \nonumber \\
\frac{\partial^2}{\partial T^2}\rho_{T>T_U}(T\to T_U)&\neq &\frac{\partial^2}{\partial T^2}\rho_{T<T_U}(T\to T_U)\,,
\label{rho1 rho2 compare}
\end{eqnarray}
in particular,
\begin{eqnarray}
\frac{\partial^2}{\partial T^2}\rho_{T>T_U}(T\to T_U)&=&\frac{13}{30}a^2\,, \quad \frac{\partial^2}{\partial T^2}\rho_{T<T_U}(T\to T_U)=\frac{73}{30}a^2\,, \nonumber \\
\frac{\partial^2}{\partial T^2}\rho_{T<T_U}-\frac{\partial^2}{\partial T^2}\rho_{T>T_U}&=&2 a^2
\,.
\label{disc1}
\end{eqnarray}
Thus, we see that crossing of the pole leads to an instability in the energy density, which manifests itself in the discontinuity of the second order derivative $\frac{\partial^2 \rho}{\partial T^2}$ at the Unruh temperature. It is also easy to see from  (\ref{rho1 rho2 compare}) that near, but below $T_U$, the energy density is negative \cite{Becattini:2017ljh, Florkowski:2018myy}, which also indicates instability in the system .

The order of the derivative $\frac{\partial^d \rho}{\partial T^d}$, in which discontinuity occurs, turns out to be related to the pole order $k$ in  (\ref{Is1 int eta}) (where $k=2$) and the order of polynomial weight $n$ in integrals (\ref{Is1Is2}) as follows
\begin{eqnarray}
d=n-k+2\,.
\label{order}
\end{eqnarray}
The derivation of formula (\ref{order}) is given in Appendix {\ref{Sec:order}. From (\ref{order}), the appearance of a discontinuity in the second order derivative in (\ref{rho1 rho2 compare}) is obvious as minimal $n$ equals 2 and $k=2$ in (\ref{main}).

It is clear that as the temperature decreases further, the integration contour $I_{s1}$ (and $I_{s2}$) cross the pole $\eta=-1$ again Appendix \ref{Sec:beyond}.

%=================================================
\section{Discussion}
\label{Sec:disc}
%=================================================

Let us turn to possible applications of our results to heavy-ion collisions.
It is known that the final state can be described as thermal hadronic excitations over the standard Minkowski vacuum. 
Combining this well-known fact with the instability for $T < T_U$ described in the previous sections we come to the key 
point that the observed hadronic spectrum could appear as a result of {\it decay} of this unstable state.

While the relation of hadronization to Unruh effect was first introduced in papers \cite{Kharzeev:2006aj,Castorina:2007eb, Becattini:2008tx},
the role of intermediate unstable state had not been discussed, to our best knowledge. 

The analytic continuation of our basic result (\ref{main}) allows us to evaluate the energy density down to $T=0$.
The resulting tricky oscillating behaviour (see Appendix \ref{Sec:beyond}), when applied to hadronization, may lead to appearance of 
a sort of mixed phase. However, our derivation corresponds to equilibrium so that only the region $ (T_U - T) \ll T_U$ can be described rigorously.

We note that similar conclusions have been reached in the recent  
field-theoretical analysis \cite{Pimentel:2018fuy}, where the instability of vacuum in strong external fields was considered. 
Cases of a scalar field with an external potential, an electric field \cite{Gies:2015hia}, and a gravitational field were addressed. 
The key observations concern the fields being above the critical values allowing the unsuppressed pair production. It turns out that close to the threshold values the reliable calculations of critical exponents and vacuum decay rates are possible. 

As in \cite{Pimentel:2018fuy}, we show that violation of the classical geometric constraint is accompanied by instability at the quantum level, and the transition through this instability is smooth.  However, despite of the similarity of results between quantum-statistical and field-theoretical examples, there are certain differences in technical details. In particular, in \cite{Pimentel:2018fuy} quantum instability is due to imaginary part to the effective action. In thermodynamics, on the other hand, energy density becomes negative, but remains real and the discontinuity manifests itself at jump of the second-order derivative (\ref{disc1}). Also energy density as the function of acceleration appears to be even for $T>T_U $, but becomes odd below $ T_U $.

Note, however, that the validity of the results based on the analytical continuation  of the Eq.~(\ref{main}) deeply into the region $ T<T_U $ is questionable. Indeed, Eq.~(\ref{main}) originally refers to equilibrium. However, as it is discussed above, the states with $ T<T_U $ are unstable. Therefore, the results based on the analytical continuation of Eq.~(\ref{main}) to $ T<T_U $ can be trusted only as far as the effect of instability is small, or close to the point $ T=T_U $.

%=================================================
\section{Conclusions}
\label{Sec:Concl}
%=================================================

In the first part of the present paper, we showed the exact correspondence of the fermion energy-momentum tensor calculated in the flat space, described by the Minkowski metric, based on the Zubarev quantum-statistical density operator and based on the heat kernel in a space with a conical singularity. In particular, this is manifested in the exact correspondence of the formulas (\ref{perturb}) and (\ref{energ Dow1}) (up to a factor of $1/2$, associated with a different number of degrees of freedom).

The found correspondence establish polynomiality of the Eq.~(\ref{perturb}) and absence of higher order corrections $ a^n,\, n=6,8... $. On the other hand, we have shown that the polynomiality of the Sommerfeld integrals, which describe the energy density of the accelerated fermion gas, can be easily found by transforming them into contour integrals in the complex plane.

Let's notice that polynomial Sommerfeld integrals exist for any integer dimension of the integrand. The lowest dimensional examples are known to be related to quantum anomalies \cite{Stone:2018zel}. In general case Sommerfeld integrals are expected to allow to obtain exact one-loop results. It is not ruled out that this is more general phenomenon, than anomalies.

Further, we show that at Unruh temperature several processes take place simultaneously. From the point of view of a space with a conical singularity, the angular deficit of the cone reaches its limiting value and the cone turns into a plane. This behaviour makes the analysis of region $ T<T_U $ problematic within the framework of the conical singularity approach.

At the quantum-statistical level, the integration contour crosses the pole of the thermodynamic distribution in the complex momentum plane, as a result of which the second derivative of the energy density has discontinuity at $ T=T_U $. Moreover, the consideration of acceleration as an imaginary part of the chemical potential makes it possible to construct an analytic continuation into region $ T<T_U $.

As the result we show, that at $ T<T_U $, odd terms in acceleration appear in the energy density, and also the energy density becomes a negative. It turns out, however, that the result differs from the perturbative calculation, that is, nonperturbative effects become significant.

The described features of states at $ T<T_U $ allow us to interpret them as unstable states. By analogy with the phenomenon of superradiance \cite{Brito:2015oca}, the decay of these unstable states should be accompanied by particle production, which may have applications in heavy-ion physics and explain thermalization and hadronization, expanding the approach using the Unruh effect to describe the thermal hadronic spectrum \cite{Kharzeev:2006aj,Castorina:2007eb, Becattini:2008tx}.

The described instability is similar to the results of the analysis of \cite{Pimentel:2018fuy, Gies:2015hia}, where non-thermodynamic instability of the vacuum was discussed. Like the analysis in \cite{Pimentel:2018fuy, Gies:2015hia}, we have a violation of the constraint resulting from geometry at the classical level, which is accompanied by quantum instability. As in \cite{Pimentel:2018fuy, Gies:2015hia}, we constructed an analytic continuation into the region of instability and show that the transition through instability is smooth.

%=================================================
{\bf Acknowledgements}
%=================================================

We are grateful to F. Becattini, M. Bordag, D. Fursaev and I. Pirozhenko for useful discussions and comments.
The work was supported by Russian Science Foundation
Grant No 16-12-10059.

\appendix

%=============================================
\section{The order of the derivative with discontinuity}
\label{Sec:order}
%=============================================

The purpose of this appendix is to derive Eq.~(\ref{order}) for the order of the derivative in which instability occurs at the Unruh temperature. To do this, let's consider the integral of the form
\begin{eqnarray}
I_{s}=\frac{T^4}{\pi^2} y^l \Big(\int_0^{\infty}\frac{x^n}{e^{x+iy}+1}+(-1)^{n+1}\int_0^{\infty}\frac{x^n}{e^{x-iy}+1}\Big)
\,,
\label{Is ord}
\end{eqnarray}
where $y=\frac{a}{2T}$. Eq.~(\ref{Is ord}) is a special case of  (\ref{Is1Is2}) for polynomial weight functions. For $l=0, n=3$ and $l=1, n=2$, one obtains integrals from  (\ref{rho with Is1 Is2}). As described in Sec.~\ref{Sec:inst}, at $T<\frac{a}{2\pi}$, there is an additional contribution to this integral due to the crossing of the pole. From  (\ref{Is1 final}) it follows that  $\Delta I_s=I_s^{T<T_U}-I_s^{T>T_U}$ has the following form
\begin{eqnarray}
\Delta I_{s}=-\frac{T^4 y^l}{\pi^2}2\pi i\, \mathrm{Res}_{\eta=-1}\frac{\eta^D}{(\eta+1)^k}\frac{(-iy)^{n+1}}{n+1}
\,,
\label{dIs}
\end{eqnarray}
where compared to  (\ref{Is1 final}) we do not fix the order of the pole $k$.  Finding a residue, we get
\begin{eqnarray}
\Delta I_{s}&=&\frac{(-1)^{k}2 i T^4 y^l }{\pi (k-1)!  (n+1)}D (D -1)...(D-(k-2)) e^{i\pi D}(-iy)^{n+1} \nonumber \\
&=&\frac{(-1)^{k}2 i T^4 y^l }{\pi (k-1)!  (n+1)}(D^{k-1}+...)\big(i(\pi- y)\big)^{n+1}\nonumber \\
&=&
\frac{(-1)^k 2 i T^4 y^l n!}{\pi (k-1)!(n-k+2)!}\Big(\big(i(\pi-y)\big)^{n-k+2}+...\Big)
\,,
\label{dIs1}
\end{eqnarray}
where we used the property of the translation operator $e^{i\pi D}(-i y)^{n+1}=(i(\pi- y))^{n+1}$ and in the brackets hold the term of the highest order in $D$, and then of the lowest order in $i(\pi-y)$. The derivative of order $d$ in temperature $\frac{\partial^d}{\partial T^d}$ from $\Delta I_{s}$ at the point $T=\frac{a}{2\pi}$ will be
\begin{eqnarray}
\frac{\partial^d}{\partial T^d}\Delta I_{s}\Big|_{T=\frac{a}{2\pi}}=
\frac{(-1)^k 2 i n! a^l}{\pi (k-1)!(n-k+2)!2^l}\frac{\partial^d}{\partial T^d}
\Big(T^{4-l} \Big[\big(i(\pi-\frac{a}{2T})\big)^{n-k+2}+...\Big]\Big)\Big|_{T=\frac{a}{2\pi}}\,.
\label{ddIs}
\end{eqnarray}
It is obvious that $\frac{\partial^d}{\partial T^d}\Delta I_{s}\Big|_{T_U}$ is not zero if
\begin{eqnarray}
d\geq n-k+2
\,.
\label{mnk}
\end{eqnarray}
Thus, instability at $T_U$ appears, starting at $d=n-k+2$, as indicated in  (\ref{order}). It is also easy to find the corresponding derivative discontinuity
\begin{eqnarray}
\frac{\partial^d}{\partial T^d}\Delta I_{s}\Big|_{T_U}=
\frac{i^{n+k-1} 2^{n-k-1} \pi^{2n-2k+l-1} n!}{(k-1)!}a^{k-n+2}
\,.
\label{jump}
\end{eqnarray}
One can easily derive jump (\ref{disc1}) from (\ref{jump}).

%=============================================
\section{Instabilities arising from repeated pole crossings}
\label{Sec:beyond}
%=============================================

In this appendix, we show that when $T<T_U$, a series of instabilities arise, similar to those discussed in Sec.~\ref{Sec:inst}, due to repeated crossing of the pole by the integration contour. Let the domains  $\pi (2n-1)<y<\pi (2n+1),\, n=0,1,2...$ correspond to the integral  $I^{(n)}_{s1}$ (and analogically for $I^{(n)}_{s2}$). Then (\ref{Is1 final}), (\ref{Is2 final}) define $I^{(0)}_{s1}$ and $I^{(0)}_{s2}$. Taking into account that each time the pole is crossed, the integral over the entire contour changes by the value of the residue at the pole, we can write the recurrence equation (we consider $I^{(n)}_{s1}$)
\begin{eqnarray}
-I^{(n+1)}_{s1}+I^{(n)}_{s1}=2 \pi i D e^{i\pi (2n+1)(D-1)} F(-iy)\,.
\label{recurr}
\end{eqnarray}
Eq.~(\ref{recurr}) can be easily solved, and it leads to a finite geometric progression
\begin{eqnarray}
I^{(n)}_{s1}=I^{(0)}_{s1}+2 \pi i D \sum_{k=0}^{k=n-1} e^{i\pi (2k+1)D} F(-iy)\,.
\label{recurr1}
\end{eqnarray}
Taking into account the zero term  (\ref{Is1 final}), and then summing the terms of geometric progression, we obtain the n-th term
\begin{eqnarray}
I^{(n)}_{s1}&=&e^{2i n \pi D}\frac{\pi D}{\sin(\pi D)} F(-iy)=\Big(1+2 i n \pi D+(\frac{1}{6}-2n^2)(\pi D)^2+\frac{i}{3} (n-4 n^3) (\pi D)^3\nonumber \\
&&+\frac{1}{360}(7-120 n^2+240n^4)(\pi D)^4+\mathcal{O}\big((\pi D)^5\big)\Big)F(-iy)
\,.
\label{Is1n final}
\end{eqnarray}
Calculating now the energy density, and taking into account that $n=\Big\lfloor\frac{1}{2}+\frac{a}{4\pi T} \Big\rfloor$, we get
\begin{eqnarray}
\rho &=&\frac{7 \pi ^2 T^4}{60}+\frac{T^2 a^2}{24} -\frac{17a^4}{960\pi^2}+
\Big(\frac{\pi T^3 a}{3}+\frac{T a^3}{4 \pi}\Big)\Big\lfloor\frac{1}{2}+\frac{a}{4\pi T} \Big\rfloor \nonumber \\
&&-
\Big(\frac{T^2 a^2}{2}+2\pi^2 T^4\Big) \Big\lfloor\frac{1}{2}+\frac{a}{4\pi T} \Big\rfloor^2-
\frac{4 \pi T^3 a}{3}\Big\lfloor\frac{1}{2}+\frac{a}{4\pi T} \Big\rfloor^3
+4 \pi^2 T^4 \Big\lfloor\frac{1}{2}+\frac{a}{4\pi T} \Big\rfloor^4
\,.
\label{rho gen}
\end{eqnarray}
Thus, we have reproduced Eq.~(3.4) from \cite{Prokhorov:2019cik}, previously obtained on the basis of the properties of polylogarithms. It contains instabilities at  $T=T_U/(2k+1)\,,\,k=0,1..$, which lead at each point to the discontinuities of the second derivative $\frac{\partial^2 \rho}{\partial T^2}$.

Eq.~(\ref{rho gen}) allows formally to obtain the energy density of the accelerated gas for arbitrarily low temperatures: the corresponding plot was shown in   \cite{Prokhorov:2019cik} in Fig.~2. At the same time, as the temperature decreases, it turns out that all the coefficients at $T^4, T^2 a^2, T^3 a ...$ in  (\ref{rho gen}) begin to change (except for $a^4$) and can become arbitrarily large in absolute value for big values of index $n$.

%===========================


\begin{thebibliography}{99}
%===========================

%\cite{Kharzeev:2012ph}
\bibitem{Kharzeev:2012ph}
  D.~E.~Kharzeev, K.~Landsteiner, A.~Schmitt and H.~U.~Yee,
  ``'Strongly interacting matter in magnetic fields': an overview,''
  Lect.\ Notes Phys.\  {\bf 871} (2013) doi:10.1007/978-3-642-37305-3\_1
  [arXiv:1211.6245 [hep-ph]].
  %%CITATION = doi:10.1007/978-3-642-37305-3_1;%%
  %144 citations counted in INSPIRE as of 04 Jul 2017

%\cite{Florkowski:2018fap}
\bibitem{Florkowski:2018fap}
  W.~Florkowski and R.~Ryblewski,
  ``Hydrodynamics with spin --- pseudo-gauge transformations, semi-classical expansion, and Pauli-Lubanski vector,''
  arXiv:1811.04409 [nucl-th].
  %%CITATION = ARXIV:1811.04409;%%
  %12 citations counted in INSPIRE as of 08 Jun 2019

%\cite{Becattini:2019ntv}
\bibitem{Becattini:2019ntv}
  F.~Becattini, G.~Cao and E.~Speranza,
  ``Polarization transfer in hyperon decays and its effect in relativistic nuclear collisions,''
  arXiv:1905.03123 [nucl-th].
  %%CITATION = ARXIV:1905.03123;%%
  %2 citations counted in INSPIRE as of 08 Jun 2019

%\cite{Rogachevsky:2010ys}
\bibitem{Rogachevsky:2010ys}
  O.~Rogachevsky, A.~Sorin and O.~Teryaev,
  ``Chiral vortaic effect and neutron asymmetries in heavy-ion collisions,''
  Phys.\ Rev.\ C {\bf 82} (2010) 054910
  doi:10.1103/PhysRevC.82.054910
  [arXiv:1006.1331 [hep-ph]].
  %%CITATION = doi:10.1103/PhysRevC.82.054910;%%
  %31 citations counted in INSPIRE as of 22 Jun 2017

%\cite{Baznat:2017jfj}
\bibitem{Baznat:2017jfj}
  M.~Baznat, K.~Gudima, A.~Sorin and O.~Teryaev,
  ``Hyperon polarization in heavy-ion collisions and holographic gravitational anomaly,''
  Phys.\ Rev.\ C {\bf 97}, no. 4, 041902 (2018)
  doi:10.1103/PhysRevC.97.041902
  [arXiv:1701.00923 [nucl-th]].
  %%CITATION = doi:10.1103/PhysRevC.97.041902;%%
  %18 citations counted in INSPIRE as of 08 Jun 2019

%\cite{Jiang:2016wvv}
\bibitem{Jiang:2016wvv}
  Y.~Jiang and J.~Liao,
  ``Pairing Phase Transitions of Matter under Rotation,''
  Phys.\ Rev.\ Lett.\  {\bf 117}, no. 19, 192302 (2016)
  doi:10.1103/PhysRevLett.117.192302
  [arXiv:1606.03808 [hep-ph]].
  %%CITATION = doi:10.1103/PhysRevLett.117.192302;%%
  %23 citations counted in INSPIRE as of 08 Jun 2019

%\cite{Chernodub:2017ref}
\bibitem{Chernodub:2017ref}
  M.~N.~Chernodub and S.~Gongyo,
  ``Effects of rotation and boundaries on chiral symmetry breaking of relativistic fermions,''
  Phys.\ Rev.\ D {\bf 95}, no. 9, 096006 (2017)
  doi:10.1103/PhysRevD.95.096006
  [arXiv:1702.08266 [hep-th]].
  %%CITATION = doi:10.1103/PhysRevD.95.096006;%%
  %25 citations counted in INSPIRE as of 08 Jun 2019

%\cite{Wang:2018sur}
\bibitem{Wang:2018sur}
  X.~Wang, M.~Wei, Z.~Li and M.~Huang,
  ``Quark matter under rotation in the NJL model with vector interaction,''
  Phys.\ Rev.\ D {\bf 99}, no. 1, 016018 (2019)
  doi:10.1103/PhysRevD.99.016018
  [arXiv:1808.01931 [hep-ph]].
  %%CITATION = doi:10.1103/PhysRevD.99.016018;%%
  %6 citations counted in INSPIRE as of 08 Jun 2019

%\cite{Prokhorov:2019cik}
\bibitem{Prokhorov:2019cik}
  G.~Y.~Prokhorov, O.~V.~Teryaev and V.~I.~Zakharov,
  ``Unruh effect for fermions from the Zubarev density operator,''
  Phys.\ Rev.\ D {\bf 99}, no. 7, 071901 (2019)
  doi:10.1103/PhysRevD.99.071901
  [arXiv:1903.09697 [hep-th]].
  %%CITATION = doi:10.1103/PhysRevD.99.071901;%%

%\cite{Buzzegoli:2017cqy}
\bibitem{Buzzegoli:2017cqy}
  M.~Buzzegoli, E.~Grossi and F.~Becattini,
  ``General equilibrium second-order hydrodynamic coefficients for free quantum fields,''
  JHEP {\bf 1710} (2017) 091
  doi:10.1007/JHEP10(2017)091
  [arXiv:1704.02808 [hep-th]].
  %%CITATION = doi:10.1007/JHEP10(2017)091;%%
  %2 citations counted in INSPIRE as of 01 Nov 2017


%\cite{Becattini:2017ljh}
\bibitem{Becattini:2017ljh}
  F.~Becattini,
  ``Thermodynamic equilibrium with acceleration and the Unruh effect,''
  Phys.\ Rev.\ D {\bf 97}, no. 8, 085013 (2018)
  doi:10.1103/PhysRevD.97.085013
  [arXiv:1712.08031 [gr-qc]].
  %%CITATION = doi:10.1103/PhysRevD.97.085013;%%
  %9 citations counted in INSPIRE as of 12 Mar 2019

%\cite{Unruh:1976db}
\bibitem{Unruh:1976db}
  W.~G.~Unruh,
  ``Notes on black hole evaporation,''
  Phys.\ Rev.\ D {\bf 14}, 870 (1976).
  doi:10.1103/PhysRevD.14.870.
  %%CITATION = doi:10.1103/PhysRevD.14.870;%%
  %2784 citations counted in INSPIRE as of 12 Mar 2019

%\cite{Dowker:1994fi}
\bibitem{Dowker:1994fi}
  J.~S.~Dowker,
  ``Remarks on geometric entropy,''
  Class.\ Quant.\ Grav.\  {\bf 11}, L55 (1994)
  doi:10.1088/0264-9381/11/4/001
  [hep-th/9401159].
  %%CITATION = doi:10.1088/0264-9381/11/4/001;%%
  %76 citations counted in INSPIRE as of 11 Oct 2019

%\cite{Dowker:1987pk}
\bibitem{Dowker:1987pk}
  J.~S.~Dowker,
  ``Vacuum Averages for Arbitrary Spin Around a Cosmic String,''
  Phys.\ Rev.\ D {\bf 36}, 3742 (1987).
  doi:10.1103/PhysRevD.36.3742
  %%CITATION = doi:10.1103/PhysRevD.36.3742;%%
  %112 citations counted in INSPIRE as of 11 Oct 2019

%\cite{Prokhorov:2018qhq}
\bibitem{Prokhorov:2018qhq}
  G.~Prokhorov, O.~Teryaev and V.~Zakharov,
  ``Axial current in rotating and accelerating medium,''
  Phys.\ Rev.\ D {\bf 98}, no. 7, 071901 (2018)
  doi:10.1103/PhysRevD.98.071901
  [arXiv:1805.12029 [hep-th]].
  %%CITATION = doi:10.1103/PhysRevD.98.071901;%%
  %6 citations counted in INSPIRE as of 30 Nov 2018


%\cite{Brito:2015oca}
\bibitem{Brito:2015oca}
  R.~Brito, V.~Cardoso and P.~Pani,
  ``Superradiance : Energy Extraction, Black-Hole Bombs and Implications for Astrophysics and Particle Physics,''
  Lect.\ Notes Phys.\  {\bf 906}, pp.1 (2015)
  doi:10.1007/978-3-319-19000-6
  [arXiv:1501.06570 [gr-qc]].
  %%CITATION = doi:10.1007/978-3-319-19000-6;%%
  %231 citations counted in INSPIRE as of 24 Oct 2019

%\cite{Pimentel:2018fuy}
\bibitem{Pimentel:2018fuy}
  G.~L.~Pimentel, A.~M.~Polyakov and G.~M.~Tarnopolsky,
  ``Vacua on the Brink of Decay,''
  Rev.\ Math.\ Phys.\  {\bf 30}, no. 07, 1840013 (2018)
  doi:10.1142/S0129055X18400135, 10.1142/9789813233867\_0020
  [arXiv:1803.09168 [hep-th]].
  %%CITATION = doi:10.1142/S0129055X18400135, 10.1142/9789813233867_0020;%%
  %5 citations counted in INSPIRE as of 11 Oct 2019

%\cite{Gies:2015hia}
\bibitem{Gies:2015hia}
  H.~Gies and G.~Torgrimsson,
  ``Critical Schwinger pair production,''
  Phys.\ Rev.\ Lett.\  {\bf 116}, no. 9, 090406 (2016)
  doi:10.1103/PhysRevLett.116.090406
  [arXiv:1507.07802 [hep-ph]].
  %%CITATION = doi:10.1103/PhysRevLett.116.090406;%%
  %22 citations counted in INSPIRE as of 26 Oct 2019

%\cite{Kharzeev:2006aj}
\bibitem{Kharzeev:2006aj}
  D.~Kharzeev,
  ``Hawking-Unruh phenomenon in the parton language,''
  Eur.\ Phys.\ J.\ A {\bf 29}, 83 (2006).
  doi:10.1140/epja/i2005-10302-1
  %%CITATION = doi:10.1140/epja/i2005-10302-1;%%
  %6 citations counted in INSPIRE as of 26 Oct 2019


 %\cite{Castorina:2007eb}
\bibitem{Castorina:2007eb}
  P.~Castorina, D.~Kharzeev and H.~Satz,
  ``Thermal Hadronization and Hawking-Unruh Radiation in QCD,''
  Eur.\ Phys.\ J.\ C {\bf 52}, 187 (2007)
  doi:10.1140/epjc/s10052-007-0368-6
  [arXiv:0704.1426 [hep-ph]].
  %%CITATION = doi:10.1140/epjc/s10052-007-0368-6;%%
  %134 citations counted in INSPIRE as of 04 Feb 2019

%\cite{Becattini:2008tx}
\bibitem{Becattini:2008tx}
  F.~Becattini, P.~Castorina, J.~Manninen and H.~Satz,
  ``The Thermal Production of Strange and Non-Strange Hadrons in e+ e- Collisions,''
  Eur.\ Phys.\ J.\ C {\bf 56}, 493 (2008)
  doi:10.1140/epjc/s10052-008-0671-x
  [arXiv:0805.0964 [hep-ph]].
  %%CITATION = doi:10.1140/epjc/s10052-008-0671-x;%%
  %79 citations counted in INSPIRE as of 06 Feb 2019

%\cite{Zubarev}
\bibitem{Zubarev}
D. N. Zubarev, A. V. Prozorkevich, S. A. Smolyanskii, "Derivation of nonlinear generalized equations of quantum relativistic hydrodynamics", TMF, 40:3 (1979), 394-407; Theoret. and Math. Phys., 40:3 (1979), 821-831.

%\cite{Weert}
\bibitem{Weert}
G. Van Weert,
"Maximum entropy principle and relativistic hydrodynamics",
Ch. Annals Phys. Volume 140, Issue 1, (1982), 133-162.

%\cite{Florkowski:2018myy}
\bibitem{Florkowski:2018myy}
  W.~Florkowski, E.~Speranza and F.~Becattini,
  ``Perfect-fluid hydrodynamics with constant acceleration along the stream lines and spin polarization,''
  Acta Phys.\ Polon.\ B {\bf 49}, 1409 (2018)
  doi:10.5506/APhysPolB.49.1409
  [arXiv:1803.11098 [nucl-th]].
  %%CITATION = doi:10.5506/APhysPolB.49.1409;%%
  %18 citations counted in INSPIRE as of 12 Mar 2019

%\cite{Becattini:2013fla}
\bibitem{Becattini:2013fla}
  F.~Becattini, V.~Chandra, L.~Del Zanna and E.~Grossi,
  ``Relativistic distribution function for particles with spin at local thermodynamical equilibrium,''
  Annals Phys.\  {\bf 338} (2013) 32
  doi:10.1016/j.aop.2013.07.004
  [arXiv:1303.3431 [nucl-th]].
  %%CITATION = doi:10.1016/j.aop.2013.07.004;%%
  %39 citations counted in INSPIRE as of 22 Jun 2017

%\cite{Fulling:2018lez}
\bibitem{Fulling:2018lez}
  S.~A.~Fulling and J.~H.~Wilson,
  ``The Equivalence Principle at Work in Radiation from Unaccelerated Atoms and Mirrors,''
  Phys.\ Scripta {\bf 94}, no. 1, 014004 (2019)
  doi:10.1088/1402-4896/aaecaa
  [arXiv:1805.01013 [quant-ph]].
  %%CITATION = doi:10.1088/1402-4896/aaecaa;%%
  %4 citations counted in INSPIRE as of 08 Jun 2019

%\cite{Stone:2018zel}
\bibitem{Stone:2018zel}
  M.~Stone and J.~Kim,
  ``Mixed Anomalies: Chiral Vortical Effect and the Sommerfeld Expansion,''
  Phys.\ Rev.\ D {\bf 98}, no. 2, 025012 (2018)
  doi:10.1103/PhysRevD.98.025012
  [arXiv:1804.08668 [cond-mat.mes-hall]].
  %%CITATION = doi:10.1103/PhysRevD.98.025012;%%
  %6 citations counted in INSPIRE as of 30 Nov 2018

\bibitem{Blankenbecler}
R. Blankenbecler,
 ``Integrals over the Fermi Function,''
1957 Am. J. Phys. 25 279–80.

\bibitem{Sprung}
D. W. L. Sprung and J. Martorell,
``The symmetrized Fermi function and its transforms,''
 1997 J. Phys. A: Math. Gen. 30 6525.

\bibitem{Burov}
V. V. Burov , F. A. Ivanyuk and B. D. Konstantinov,
``Effect of the Nuclear Charge Density Oscillations in Elastic Scattering of Electrons,''
 1975 Yad. Fiz. 22 1142–5.

%\cite{Gribov:2009zz}
\bibitem{Gribov:2009zz}
  V.~N.~Gribov, Y.~L.~Dokshitzer and J.~Nyiri,
  ``Strong interactions of hadrons at high emnergies: Gribov lectures on,''
  Camb.\ Monogr.\ Part.\ Phys.\ Nucl.\ Phys.\ Cosmol.\  {\bf 27} (2012).
  %%CITATION = CMPCE,27,;%%
  %8 citations counted in INSPIRE as of 21 Oct 2019

%\cite{Fursaev:1993qk}
\bibitem{Fursaev:1993qk}
  D.~V.~Fursaev,
  ``The Heat kernel expansion on a cone and quantum fields near cosmic strings,''
  Class.\ Quant.\ Grav.\  {\bf 11}, 1431 (1994)
  doi:10.1088/0264-9381/11/6/008
  [hep-th/9309050].
  %%CITATION = doi:10.1088/0264-9381/11/6/008;%%
  %61 citations counted in INSPIRE as of 14 Oct 2019

%\cite{Mertens:2015ola}
\bibitem{Mertens:2015ola}
  T.~G.~Mertens,
  ``Hagedorn String Thermodynamics in Curved Spacetimes and near Black Hole Horizons,''
  arXiv:1506.07798 [hep-th].
  %%CITATION = ARXIV:1506.07798;%%
  %2 citations counted in INSPIRE as of 21 Oct 2019

%\cite{Adler:1969er}
\bibitem{Adler:1969er}
  S.~L.~Adler and W.~A.~Bardeen,
  ``Absence of higher order corrections in the anomalous axial vector divergence equation,''
  Phys.\ Rev.\  {\bf 182}, 1517 (1969).
  doi:10.1103/PhysRev.182.1517
  %%CITATION = doi:10.1103/PhysRev.182.1517;%%
  %1064 citations counted in INSPIRE as of 21 Oct 2019

%\cite{Sadofyev:2010is}
\bibitem{Sadofyev:2010is}
  A.~V.~Sadofyev, V.~I.~Shevchenko and V.~I.~Zakharov,
  ``Notes on chiral hydrodynamics within effective theory approach,''
  Phys.\ Rev.\ D {\bf 83}, 105025 (2011)
  doi:10.1103/PhysRevD.83.105025
  [arXiv:1012.1958 [hep-th]].
  %%CITATION = doi:10.1103/PhysRevD.83.105025;%%
  %92 citations counted in INSPIRE as of 03 Jul 2019

%\cite{Son:2009tf}
\bibitem{Son:2009tf}
  D.~T.~Son and P.~Surowka,
  ``Hydrodynamics with Triangle Anomalies,''
  Phys.\ Rev.\ Lett.\  {\bf 103}, 191601 (2009)
  doi:10.1103/PhysRevLett.103.191601
  [arXiv:0906.5044 [hep-th]].
  %%CITATION = doi:10.1103/PhysRevLett.103.191601;%%
  %604 citations counted in INSPIRE as of 03 Jul 2019

\end{thebibliography}
\end{document}